\documentclass[10.5pt,onecolumn,letter]{IEEEtran}
\usepackage[margin=1in]{geometry}
\usepackage[english]{babel}
\usepackage[linesnumbered, ruled,vlined]{algorithm2e}
\usepackage{graphicx,subfigure}
\usepackage{amsmath,amssymb,latexsym}
\usepackage{mathtools, cuted}
\usepackage{amsfonts}
\usepackage{tikz}
\usepackage{lipsum}
\usepackage{color,setspace}
\usepackage{epstopdf}
\usepackage{wrapfig}
\graphicspath{{./fig/}}
\begin{document}
\title{\large \bf Enabling Millimeter-Wave 5G Networks for Massive IoT Applications}
\author{By Biswa~P.~S.~Sahoo, Ching-Chun Chou, Chung-Wei Weng, and  Hung-Yu~Wei}

\maketitle
%
%\doublespacing
\begin{abstract}
Internet of Things is one of the most promising technology of the fifth-generation (5G) mobile broadband systems. Data-driven wireless services of 5G systems require unprecedented capacity and availability. The millimeter-wave based wireless communication technologies are expected to play an essential role in future 5G systems. In this article, we describe the three broad categories of fifth-generation services, viz., enhanced mobile broadband, ultra-reliable and low-latency communications, and massive machine-type communications. Furthermore, we introduce the potential issues of consumer devices under a unifying 5G framework. We provide the state-of-the-art overview with an emphasis on technical challenges when applying millimeter-wave (mmWave) technology to support the massive Internet of Things applications. Our discussion highlights the challenges and solutions, particularly for communication/computation requirements in consumer devices under the millimeter-wave 5G framework.
\end{abstract}

\IEEEpeerreviewmaketitle

\section{Introduction}
\begin{wrapfigure}{r}{0.6\textwidth}
\centering
\includegraphics[width=0.6\textwidth]{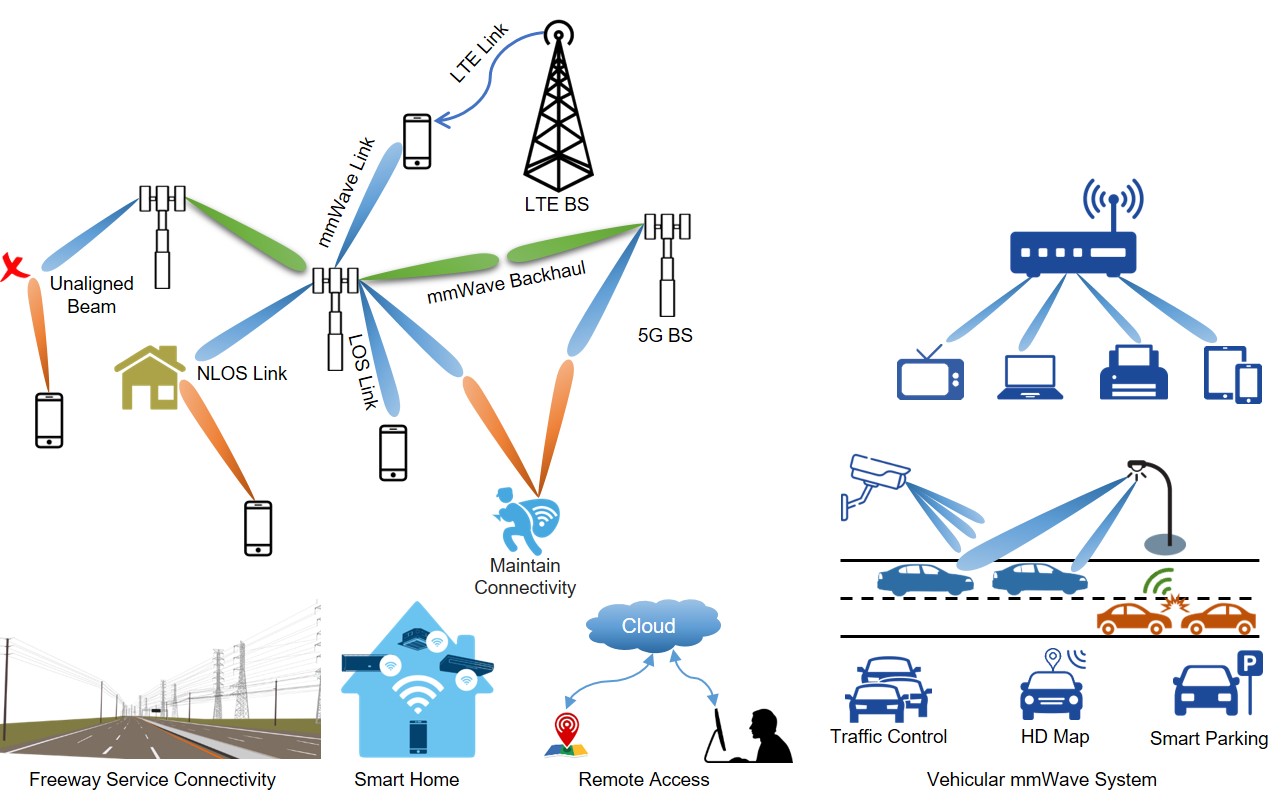}
\caption{5G millimeter-wave beamforming system for IoT applications.}
\label{fig:5gnetwork}
\end{wrapfigure}

The radio technological revolution for wireless communication has already started. In past decade the wireless communications have been revolutionized in many folds and variety of new applications has been developed. As a result, we are witnessing game-changing applications such as smart home, industrial application and control, traffic safety and control, remote surgery, and 4K/8K video broadcasting, etc. Fig. \ref{fig:5gnetwork} shows the 5G use cases in mmWave beamforming system. This promises to change the way the wireless services use to communicate and deliver. The next generation wireless technology, popularly known as fifth generation (5G), promises to deliver new levels of capability and efficiency by supporting a diverse range of applications of future which can change the way we live, work, and communicate with each other; and able to cater many consumer needs.
\par
The 3rd Generation Partnership Project (3GPP) envisioned a diverse set of usage scenarios and applications driven by massive devices, low cost, low energy, both low latency and high reliability, and very high availability. The International Telecommunication Union (ITU) has classified 5G mobile network services into three categories, viz., eMBB (enhanced Mobile Broadband), mMTC (massive Machine-Type Communications), and uRLLC (Ultra-Reliable and Low-Latency Communications)~\cite{IMTVision}. Meanwhile, the emergence of the unprecedented Internet of Things (IoT) services and applications have attracted much attention in both industry and academia.

Providing unprecedented radio access capacity to these massive IoT applications, millimeter-wave (mmWave) communications  \cite{ZPi2011}\cite{Yilmaz2016} is emerging as one of the key enablers. However, there exist multiple challenges spanning from the physical infrastructure to the communication protocol design, to enable mmWave communications to support IoT applications. As the density of wireless devices increases and requirements of all their applications changes drastically, it becomes more challenging to provide them connectivity within the stipulated amount of time to deliver the needs. With the increase of device density, the data traffic also grows exponentially, as a result, the conventional wireless technologies preliminarily show their incompetence because of limited bandwidth \cite{Kong2017}. In this context, mmWave technology shows its potential to address this problem. However, to fully exploit mmWave capability to achieve the 5G use case target, much needed to be studied.
\par
The goal of this article is to provide an overview of these challenges and discuss the most promising solutions to enable mmWave 5G networks to support massive IoT services and applications.

\section{Millimeter-Wave-Enabled IoT Systems}
The mmWave 5G wireless network is targeting to support the broad range of vertical services categorized by eMBB, mMTC and uRLLC by a unifying technical framework. We will briefly discuss the high-level requirements and challenges for these three service categories of 5G mmWave-enabled IoT applications. We will focus our discussion on the issues related to directional transmission over mmWave bands, high mobility scenarios, ultra-dense deployments, and the last but not least, communication/computing requirements to support diverse IoT applications.

\subsection{eMBB}
Following the ITU's IMT 2020 Vision \cite{IMTVision}, the need to deliver gigabit mobile broadband, otherwise known as eMBB, is one of three distinct 5G use cases. It aims to deliver up to 10-Gbps peak throughput, 1-Gbps throughput in high mobility, and up to 10,000x total network traffic. In recent past, research in this direction has already been demonstrated that it can reach such high data rates. It is anticipated that the much-discussed mmWave technology and massive multiple-input and multiple-output, or MIMO will play a critical role to achieve this outrageous target. eMBB is targeting services such as seamless ultra-high speed data access in both indoors and outdoors, high-resolution multimedia streaming on high mobility scenarios, real-time big data delivery, and AR/VR applications. Recent study, Cisco Visual Networking Index (VNI) Global Mobile Data Traffic Forecast Update \cite{Cisco2017}, shows the average consumer's smartphone alone is expected to go from consuming 1.6 GB of data per month today to close to 7 GB of data per month in 2021. In order to achieve this increase in capacity over the current 4G systems, 5G NR offers fiber-like speeds with uniform experience for both download and upload, even in challenging environment or at the cell edge. However, mmWave cellular comes with challenges which need to be addressed to achieve the target. We focus on two critical areas such as user mobility scenarios and resources for communication/computation.
\par
A recent study on IoT-Cloud supported vehicular system for sharing multi-gigabit data about the surrounding environment and real-time object recognition is discussed in \cite{Kong2017}. This system is designed for both vehicle-to-vehicle (V2V) and vehicle-to-infrastructure (V2I) communications over the mmWave band. To fully realized the vehicular mmWave system and provide these services during mobility requires lower latency and light network congestion to process the data streams and user requests in near real-time \cite{Yang2017}. In this aspect, it is a recent research trend to bring the computing capability and services to the network edges \cite{Shih2017}. Thus, combining the edge computing and cloud will work as an enabler for the mmWave 5G ecosystem to realize eMBB services and beyond. Edge computing has the strong capability for analyzing and planning for equipment responsible for communication and resource allocation. Numerous prospective challenges in accomplishing the mmWave broadband have been established and discussed in \cite{ZPi2011}. Solutions for some of those issues were also investigated. For example, to reduce the outage probability during user mobility and maintain uninterrupted service continuity in the mmWave platform, the author in \cite{ChunHan2017} proposed a joint scheduling and power allocation framework for 5G network. Thus, to achieve this goal, the control signaling, efficient allocation of communication/computation resources are the key enabler.

%\begin{figure}[!htb]
\begin{wrapfigure}{r}{0.5\textwidth}
\centering
\includegraphics[width=0.45\columnwidth]{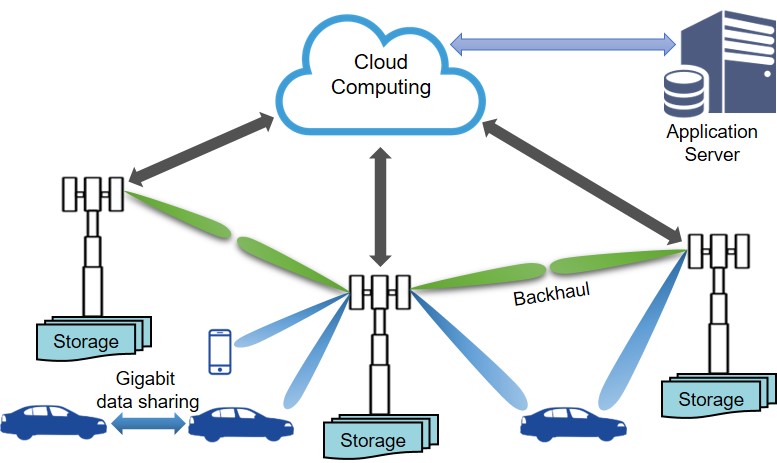}
\caption{Integrated access/backhauling to enable eMBB services.}
\label{fig:mmWCloud}
\end{wrapfigure}

In \cite{MiEdge2017}, the 5G-MiEdge - Millimeter-wave Edge - architecture, which combines mmWave access/backhauling with MEC to enable eMBB services with the efficient allocation of communication/computation resources. Motivated by  \cite{r2-164130}, we give a hybrid architecture example as shown in Fig.~\ref{fig:mmWCloud}, to enable giga-bit eMBB services. The centralized communication and storage cloud can be moved to the edge access points to provide flexible management of radio signaling and computing. Computation offloading is a key strategy to move the computation burden from resource-hungry mobile devices to more powerful fixed servers. Computation offloading and edge server will greatly benefit both the mmWave radio access and backhaul. In this hybrid architecture, the gigabit eMBB application data processing can be performed on the edge-cloud to reduce computation, and integrated access/backhauling can provide coordinated and cooperative communication-intensive. In the similar direction, the MiEdge has shown how a joint allocation of communication and computation resources can provide considerable advantages with respect to disjoint strategies.
\par
\begin{wrapfigure}{r}{0.5\textwidth}
\centering
\includegraphics[width=0.45\columnwidth]{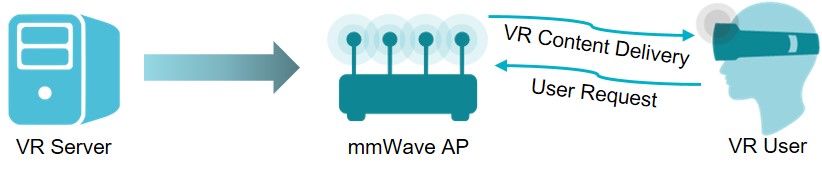}
\caption{Illustrative mmWave-based VR network.}
\label{fig:arvr}
\end{wrapfigure}

Recently, VR/AR applications have received a significant recognition. The VR applications connect to a VR data-center, called VR server (VRS), as shown in Fig.~\ref{fig:arvr}, in order to receive massive VR data. The network infrastructure should be capable of providing ultra-high-speed data communication to transmit massive VR content without lag to multiple users in crowded areas  e.g. in trains, airports or other congested indoor areas. As we discussed previously, for massive data delivery mmWave communication technologies, using steerable and directive beams, are best suited for its capability and efficiency. However, unfortunately, the VR devices will suffer from three major challenges in a mmWave based VR network such as huge energy consumption during massive data transmission; set of interference from the neighbor mmWave VR transmitter due to high density and multi-link communications. Thus, providing truly VR experiences is a key challenge. However, the challenges can be met by applying successive interference management scheme, but this, of course, requires additional complexity and it goes beyond the scope of this paper.

Fully exploiting mmWave to realize eMBB and beyond, requires significant changes at multiple layers of the protocol stack are required to reach the phenomenal objective. Recently, 3GPP has approved way forward on overall 5G~NR eMBB workplan~\cite{RP-170741}, which is expected to enabling deployments as early as 2019. 

\subsection{uRLLC}
The mmWave 5G supported IoT systems also aims to deliver critical information with high reliability, which can be defined as by a very low rate of lost packets. Such applications involve requirements on high reliability and low latency for delivering data, not necessarily with high throughput, forming thus the concept of uRLLC. For example, remote access including natural disaster, healthcare, military communications, industrial communication, as well as automotive applications which require high precision. For these applications, reliability is a critical factor over the guaranteed End-to-End (E2E) throughput. However, the very low latency can be a quite varying requirement ranging from below 1 ms E2E to multiple seconds. On the other hand, reliability could be important for an application dealing with sensitive data. For example, reliability stands as the first important factor when dealing with medical data or medical equipment in real-time. Since the applications' success rate relies on the transmission guarantee within a bounded latency limit. Thus, mmWave 5G design for IoT system should be flexible to be able to address a range of uRLLC services concerning different latency and reliability requirements.
\par
While the mmWave bands have been recognized as a promising candidate technology to support massive peak data rates for providing E2E services, realizing the ultra-low-latency while maintaining service reliability pose significant challenges for system design. The mmWave wireless links are considered as highly unreliable due to its unfavorable propagation characteristics as its transmitting beam can be obstructed very easily. Thus, many suspects, it may impact on the transmission reliability and latency. We focus on two critical areas such as core network design and mobility management.
\par
For achieving uRLLC in 5G mmWave cellular networks, we need to revisit the legacy cellular network design from its core itself. The E2E services in cellular networks suffer delay primarily from the packet routing via the core network. To reduce this E2E delay, the core network function needs to be moved closer to the edge devices \cite{YJKu2017}. Technologies such as MEC \cite{Dama2017} \cite{Corcoran2016}, Software Defined Networking~(SDN), and Fog-Radio Access Network~(F-RAN) \cite{Shih2017}  are the three recent trends which can bring the efficient computing capability to the edge of the network to meet the needs of uRLLC applications. As a result, efficient utilization of processing resources to meet the communications and computing requirements becomes an issue worthy of discussion. In addition to this, integrated access and backhaul simplify the deployment of small cells which require fully flexible resource allocation between radio access and backhaul.
\par
In order to support uRLLC in high mobility scenarios, the following requirements must be satisfied \cite{r2-164130}:
\begin{itemize}
\item {\bf Handover}: Seamlessly migrating from one radio link to another without interruption is important. The network must always ensure to provide a connection within the network or when switching to another available network without interrupting service. It should also ensure that no packets are lost or delayed during handover.
\item {\bf Redundant Links}: To provide ultra-reliability, redundant transmissions are required on both the radio link and through the network infrastructure.  The redundant links should be provided through different radio access nodes and may be provided through different access points.
\item	{\bf Transmission Schemes}: Robust transmission scheme providing simultaneous transmission across multiple radio access links must be supported. These should include both coordinated and uncoordinated transmission.
\end{itemize}

\begin{wrapfigure}{r}{0.5\textwidth}
\centering
\includegraphics[width=0.45\columnwidth]{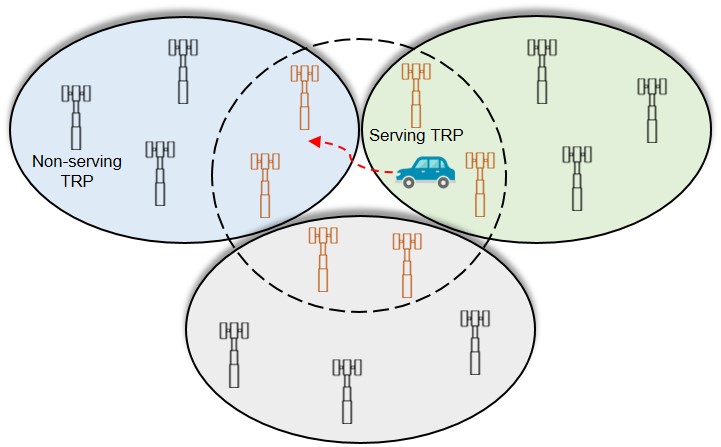}
\caption{Multi-TRP transmission for service connectivity and reliability. Serving-cluster is shown as dotted circle. TRP: Transmission Reception Point}
\label{fig:multitrp}
\end{wrapfigure}

As we mentioned previously, one of the major diminishing factors of mmWave communications is that the unreliable channel, depending on the intensity of physical blocking. Thus, the most effective strategy to counteract blocking is to enable multi-link communications via multiple radio access links. By doing so, the user can transmit/receive towards/from multiple radio access points, also known as Transmission Reception Point or multi-TRP, depending on their status. Here, inspired by \cite{r2-164130}, we give an example multiple TRP (multi-TRP) transmission during user mobility as shown in Fig.~\ref{fig:multitrp}. With a similar scenario, \cite{ChunHan2017} proposed an integrated framework is to reduce the outage probability while maintaining seamless service continuity by serving UE with multiple-beam simultaneously.

%\begin{figure}[!htb]
\par
In mmWave high mobility scenarios and in ultra-dense deployments, it is highly likely, UE will undergo frequent handover which will impact both the reliability and latency requirements due to handover interruption time and possible radio link failure (RFL).  Thus, it is important to ensure that the user is always-on connectivity for seamless handover. One way to achieve is that the user is served by at least one TRP at all times with strong received power. However, it is not only sufficient providing the connectivity but also ensure the required application data is available at the target TRP before a handover is initiated. For example, integrating edge cloud with the radio access points could be one potential solution, as previously shown in Fig.~\ref{fig:mmWCloud}. In order to satisfy ultra-high reliability and to guard against potential RLFs, it may be necessary to use redundant links on both the radio access and through the network infrastructure. In this case, for DL and UL transmission, data is available at multiple TRPs for communication with the UE.

\subsection{mMTC}
The Global Mobile Data Traffic Forecast Update \cite{Cisco2017} predicts, there will be 11.6 billion mobile-connected devices by 2021, including Machine-to-Machine (M2M) modules$-$exceeding the world's projected population at that time (7.8 billion). Besides, by 2021, a 5G connection will generate 4.7 times more traffic than the average 4G connection. Traditionally, the cellular networks or short-range network (e.g. WLAN) have been designed to support a moderate number of devices with wide coverage and mobility support. mMTC is the basis for connectivity in IoT, which allows for infrastructure management, environmental monitoring, and healthcare applications. For example, smartphone applications for healthcare integration with fitness devices; vehicular technology: mobile network integrating with roadside units, popularly known as V2X; Smart/green building integration, etc.
\par
As the number of devices increases exponentially, the diversity in communication pattern is more visible which will change the communication and computing dynamics of the system design. However, in this scenario, among these massive number of devices, not all devices will be in an active mode for all the time. For a device to be connected to the network or not, will largely be depended on the application or service type the device uses. Hence, a large number of devices will be in different mode, e.g idle mode, transmit mode, receive mode. The main challenge in mMTC is flexible and efficient connectivity for a massive number of devices sending varying packet sizes (e.g. very short packets, long packet, burst packets. etc), which has not been done adequately in cellular systems designed for human-type communications. Thus, primary the challenges in mMTC are due to the massive number of uncoordinated connections for uplink/downlink communication. As in a massive heterogeneous device centric network, UE with diverse transmission pattern will be connected to the network with different access mechanism on the same carrier or on different carriers. To ensure a smooth communication transitions between these diverse devices, the framework should include both coordinated and uncoordinated transmission schemes. Since, the inception of the cellular network, it has been designed for the voice services or uniform services assuming synchronous uplink and downlink requirements. However, those assumptions do not hold today.
\par
In order to support the massive number of devices with diverse communication pattern and requirements, aggressive technologies breaking the barrier of mmWave transmission needed to be investigated.

\section{System Evaluation}
We demonstrate the performance of the multi-TRP mmWave framework supporting both uRLLC and eMBB applications by conducting experiments through simulations. An illustration of the network model is shown in Fig.~\ref{fig:multitrp}. The network consists of serving TRPs and non-serving TRPs, and a UE is moving through the edge of multiple cells. The UE can receive from several TRPs at which UE received signal strength above a threshold. This simultaneous reception can be achieved by proper beamforming at UE side to mitigate interference. The number of simultaneous reception links depends on the number of UE antenna panels and the number of RF chains in UEs. The figure illustrates three cells with a user performing handover while moving with varying vehicular speed. In this example, multiple cells are providing redundant downlink connectivity to the user. The redundant links help the user to maintain the service connectivity while meeting service requirements.

%\begin{figure}[!htb]
\begin{wrapfigure}{r}{0.5\textwidth}
\centering
\includegraphics[width=0.4\columnwidth]{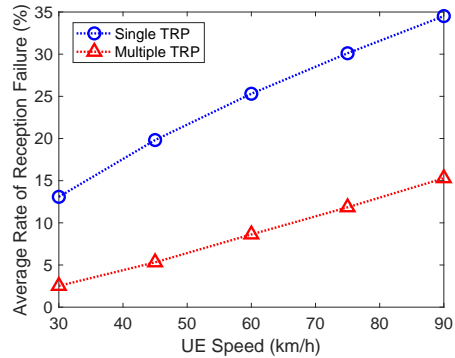}
\caption{The average rate of reception failure.}
\label{fig:outage}
\end{wrapfigure}

Fig.~\ref{fig:outage} illustrates the simulation result of the average rate of reception failure where the UE is served by multiple TRPs (e.g. two in this sample study), simultaneously. The simulation result shows the multi-TRP transmission framework always performs better than the single-TRP. The result demonstrates the redundant transmission towards UE can improve the transmission-reception failure rate, thus reliability. Further computing capability at the TRPs could help to improve the latency by reducing the round trip time. Additionally, if an added number of TRPs are further considered to serve the UE, the trends of this two curve will improve further. The multi-TRP mmWave framework also produces several interesting research directions worth for further study. 

\section{Summary and Discussion}
The mmWave band is promising a new generation of wide-area cellular networks to support massive IoT applications of future. However, will require significant changes at multiple layers of the protocol stack to reach the phenomenal objective. In this article, we have introduced the high-level requirements of three 5G use cases. In the following, we will discuss the potential issues for mmWave in consumer devices under the 5G framework for facilitating massive IoT services of the future. We categorized our discussion into four type, described below.
\par
\emph{Short Range}:  There are several mechanisms by which the consumer's end devices can access the IoT ecosystem. However, we specifically focus where the short-range wireless technologies are used. 60 GHz Wi-Fi is among the prominent access technology for IoT. The existing mmWave standards fail to fully exploit the potential advantages of short-range mmWave technology as they are designed to support limited number devices and also limited to point-to-point short-range and fixed wireless access. There is a need to redefine the whole design framework to address the IoT specific issues such as collision-aware hybrid resource allocation frameworks with on-demand control messages, the advantages of a collision notification message, and the potential of multihop communication to provide reliable mmWave connections.

\emph{Access Technology}: 5G is composed of both low-frequency (e.g. LTE) band and ultra-high frequency (e.g. mmWave) band devices. These two carriers with different physical characteristics support different access mechanism. The different career leads to different multiplexing schemes or channel access mechanism. Thus, 5G should ensure the coexistence of multiple standards. For the case of preemption to schedule, a service over LTE in DL where the resources of it can overlap with resources of ongoing/scheduled for mmWave supported service, a mechanism to avoid serious performance degradation should be considered. Mobilizing mmWave requires a new system design with tight integration with LTE or sub-6 GHz.

\emph{Mobility}: The challenges posing in mobilizing mmWave due to the high path loss and susceptibility to blockage. As we discussed previously, the smart beamforming and beam tracking is crucial to increase coverage and minimize interference so as to realize the mmWave opportunity for mobility scenario. With both line-of-sight and non-line-of-sight coverage, we will see diverse mobility pattern or scenario such as indoor mobility, outdoor mobility, and eNB handover. To drive mmWave to support in mobility scenario intelligent beam search and tracking algorithms are key for coordinated scheduling and interference management. Seamless mobility across different mmWave access technology is also another worth discussing issue. 

\emph{Other}: Besides these discussed potential problems, mmWave supported IoT also encounter many other problems as well. Here, we will highlight few of them. Enabling mmWave directional communications requires highly complex antenna processing functions and computation capability to perform directional reference signaling and scheduling. Thus, handheld devices require high battery-powered consumption. As a result, UE power saving design would be one of the most important metrics for 5G networks to see. Recently, interest in the coexistence of multiple technologies such as WiFi and LTE are emerging. However, this scope is not limited to LTE/WiFi rather likely to become a major point of discussion in the 5G ecosystem. Besides, it will be interesting to study the new risks and security threats \cite{Puthal2017} that we will be facing in 5G networks. 
\par
The unifying 5G mmWave framework to support massive IoT applications produces several promising research directions. These above-discussed potential issues are important future research directions.

\section*{Acknowledgment}
This work was financially supported by the Ministry of Science and Technology of Taiwan under Grants MOST 105-2622-8-002-002, and sponsored by MediaTek Inc., Hsin-chu, Taiwan.

\section*{About The Authors}
\par
{\noindent \bf Biswa P. S. Sahoo}  (biswap@outlook.com) is currently pursuing the Ph.D. degree with the Graduate Institute of Electrical Engineering, National Taiwan University, Taiwan.
\par
{\noindent \bf Ching-Chun Chou} (ccchou@ntu.edu.tw) is currently a Postdoctoral Fellow with the Department of Electrical Engineering, National Taiwan University, Taiwan.
\par
{\noindent \bf Chung-Wei Weng} (chungweiweng@ntu.edu.tw) is currently a Research Assistant with the Wireless Mobile Network Laboratory, Department of Electrical Engineering, National Taiwan University, Taiwan.
\par
{\noindent \bf Hung-Yu Wei} (hywei@ntu.edu.tw) is currently a professor with the Department of Electrical Engineering and the Graduate Institute of Communication Engineering, National Taiwan University, Taiwan.

\end{document}